\documentclass[preprint,12pt]{elsarticle}

\usepackage{amssymb,amsmath,amsthm,bm}

\biboptions{sort&compress}

\begin{document}

\begin{frontmatter}



\title{Electromagnetic neutrino: a short review}



\title{Theory of ionizing neutrino-atom collisions: The role of atomic recoil}


\author{Konstantin A. Kouzakov}

\address{Department of Nuclear Physics and Quantum Theory of Collisions, Faculty of Physics, Lomonosov Moscow State University, Moscow 119991, Russia}

\ead{kouzakov@srd.sinp.msu.ru}

\author[1,2]{Alexander I. Studenikin}
\ead{studenik@srd.sinp.msu.ru}

\address[1]{Department of Theoretical Physics, Faculty of Physics, Lomonosov Moscow State University, Moscow 119991, Russia}
\address[2]{Joint Institute for Nuclear Research, Dubna 141980, Moscow Region, Russia}

\begin{abstract}
We consider theoretically ionization of an atom by neutrino impact taking
into account electromagnetic interactions predicted for massive neutrinos
by theories beyond the Standard Model. The effects of atomic recoil in
this process are estimated using the one-electron and semiclassical
approximations and are found to be unimportant unless the energy transfer
is very close to the ionization threshold. We show that the energy scale
where these effects become important is insignificant for current
experiments searching for magnetic moments of reactor antineutrinos.
%The estimates
%based on a semiclassical approximation exhibit a suppression of the cross
%section due to the atomic recoil when the energy transfer
\end{abstract}


\end{frontmatter}

% \linenumbers

\section{Introduction}
\label{}
Neutrinos are very intriguing objects in particle physics. They interact
very weakly and their masses are much smaller than those of the other
fundamental fermions (charged leptons and quarks). In the Standard Model
(SM), neutrinos are massless and have only weak interactions. However, the
observation of neutrino oscillations by many experiments implies that
neutrinos are massive and mixed. Therefore, the SM must be extended to
account for neutrino masses. In many extensions of the SM, neutrinos also
acquire electromagnetic properties through quantum loop effects (see
Refs.~\cite{giunti09,broggini12,giunti14} for detail). Hence, the
theoretical and experimental study of neutrino electromagnetic
interactions is a promising tool to search for the fundamental theory
beyond the SM (BSM).

The most theoretically studied electromagnetic properties of neutrinos are
the dipole magnetic and electric moments. The neutrino magnetic moments
expected in the minimally extended SM are very small and proportional to
the neutrino masses:
$$
\mu_\nu=3\times10^{-19}\mu_B\left(\frac{m_\nu}{\rm 1\,eV}\right),
$$
with $\mu_B=e/(2m_e)$ being the electron Bohr magneton (in units
$\hbar=c=1$), and $m_e$ is the electron mass. Any larger value of
$\mu_\nu$ can arise only from the BSM
physics~\cite{giunti09,broggini12,giunti14}. Current direct experimental
searches for a magnetic moment of the electron antineutrinos from reactors
have lowered the upper limit on its value down to
$\mu_\nu<2.9\times10^{-11}\mu_B$~\cite{beda12}. These ultra low background
experiments use germanium crystal detectors exposed to the neutrino flux
from a reactor and search for scattering events by measuring the energy
deposited by the neutrino scattering in the detector. Their sensitivity to
$\mu_\nu$ crucially depends on lowering the threshold for the energy
transfer $T$. This is because the electromagnetic contribution to the
inclusive differential cross section for the neutrino scattering on a free
electron (FE) at small energy transfer ($T\ll E_\nu$) behaves as
$ {d\sigma^{\rm FE}_{(\mu)}}/{dT}\propto{1}/{T}$~\cite{vogel89},
while that induced by weak interaction, $d\sigma^{\rm FE}_{(w)}/dT$, is
practically constant in $T$~\cite{vogel89}.

The current experiments using germanium detectors have reached threshold
values of $T$ as low as few keV, where one can expect modifications of the
FE formulas due to the binding of electrons in the germanium atoms. Our
theoretical analysis (see Ref.~\cite{kouzakov14} and references therein),
involving the WKB and Thomas-Fermi models, has shown that the so-called
stepping approximation, introduced in~\cite{kopeikin97} from an
interpretation of numerical data, works with a very good accuracy. The
stepping approximation treats the process as scattering on independent
electrons occupying atomic orbitals and suggests that the cross sections
$d\sigma_{(\mu)}/dT$ and $d\sigma_{(w)}/dT$ follow the FE behaviors down
to the ionization threshold for the orbital; and below that energy the
electron on the corresponding orbital is ``inactive'', thus producing a
sharp ``step'' in the $T$ dependence of the cross section.

To the best of our knowledge, the issue of the center-of-mass atomic
motion has remained practically unadressed so far in theoretical studies
devoted to the ionization channel of neutrino-atom collisions. Usually the
recoil effects are neglected under the assumption that the atomic nucleus
due to its large mass stays at rest during the ionization process. At the
same time, it appears that in the case of reactor antineutrinos the recoil
energy can be comparable to atomic binding energies. Thus, in this
contribution, we analyze the role that can play the center-of-mass atomic
motion in the discussed processes. In particular, we inspect how it can
affect the validity of the stepping approximation.

\section{Theoretical estimate of atomic-recoil effects}
\label{theo}
We specify the incident neutrino energy and momentum by $E_\nu$
and ${\bf p}_\nu$, respectively. The atomic target is supposed to
be initially at rest, unpolarized and in its ground state
$|0\rangle$ with the corresponding energy $E_0$. We treat the
initial and final electronic systems nonrelativistically under
conditions $T\ll m_e$ and $\alpha Z\ll1$, where $Z$ is the atomic
number and $\alpha$ is the fine-structure constant. The incident
and final neutrino states are described by the Dirac spinors
assuming $m_\nu\approx0$.

First we consider the experimentally measured single-differential
inclusive cross section $d\sigma/dT$ of atomic ionization by neutrino
impact, without accounting for atomic recoil in this process. The standard
electroweak contribution to the cross section is given by (see e.g. in
\cite{kouzakov14})
\begin{eqnarray}{d \sigma_{(w)} \over dT } &=& {G_F^{2} \over 4 \pi} \left ( 1+ 4\sin^{2} \theta_{W} + 8 \sin^4 \theta_{W} \right
)\nonumber\\&{}&\times \int_{T^{2}}^{(2E_{\nu}-T)^{2}} S(T,{q}^{2}) \,
d{q}^{2}, \label{d1sw}
\end{eqnarray}
where $G_F$ is the Fermi constant, $\theta_W$ is the Weinberg
angle, and $S(T,{q}^{2})$ is the dynamical structure factor that
can be presented as follows ($q=|{\bf q}|$):
\begin{equation}S(T,{q}^{2})=\sum_n \left | \langle n |
\rho({\bf q}) | 0 \rangle \right |^{2}  \delta (T - E_n+E_0),
\label{dsf}
\end{equation}
with $\rho({\bf q})$ being the Fourier transform of the electron density
operator and the $n$ sum running over all the atomic states $| n \rangle$
with energies $E_n$ of the electron system. The $\mu_\nu$ contribution to
the cross section can be similarly expressed in terms of the same
factor~(\ref{dsf}) as~\cite{kouzakov14}
\begin{equation} {d \sigma_{(\mu)} \over dT } = 4 \pi \alpha \mu_{\nu}^{2}
\int_{T^{2}}^{(2E_{\nu}-T)^{2}} S(T,{q}^{2})\, {d{q}^{2} \over {q}^{2}},
\label{d1s} \end{equation}
where the function $S(T,{q}^{2})$ is integrated over ${q}^{2}$ with
${q}^{-2}$, rather than a unit weight as in Eq.~(\ref{d1sw}).

The kinematical limits for ${q}^{2}$ are explicitly indicated in
Eqs.~(\ref{d1sw}) and (\ref{d1s}). At large $E_{\nu}$, typical for the
reactor antineutrinos ($E_\nu\sim1$\,MeV), the upper limit can in fact be
extended to infinity, since in the discussed here nonrelativistic limit
the range of momenta $q\sim E_{\nu}$ is indistinguishable from infinity on
the atomic scale. The lower limit can be shifted to ${q}^{2}=0$, since the
contribution from the region of ${q}^{2} < T^{2}$ can be expressed in
terms of the photoelectric cross section~\cite{mv} and is negligibly small
(at the level of below one percent in the considered range of $T$). For
this reason one can discuss, without loss of accuracy, the
momentum-transfer integrals in Eqs.~(\ref{d1sw}) and~(\ref{d1s}) running
from ${q}^{2}=0$ to ${q}^{2}=\infty$.

In the one-electron approximation and in the semiclassical limit, the
dynamical structure factor~(\ref{dsf}) acquires the form~\cite{prd11}
\begin{equation} S(T,{q}^{2})={m_e \over 2pq} \, \theta(T-I)
\, \theta(q_{+}-q)\,\theta(q-q_{-}), \label{dsfWKB}
\end{equation}
%
%%
%\begin{eqnarray} S(T,{q}^{2})&=&{m_e \over 2pq} \, \theta(T-I)
%\left [ \theta \left ( T- {{q}^{2} \over 2m_e}+{pq \over m_e} \right)
%\right.\nonumber\\&{}&\left.- \theta \left ( T- {{q}^{2} \over 2m_e}- {pq
%\over m_e} \right) \right ], \label{dsfWKB}
%\end{eqnarray}
%%
where $p$ is an average momentum of the electron bound in the
atomic orbital from which ionization takes place, $I$ is the
ionization potential for this orbital, and
$$
q_{\pm}=\sqrt{p^2+2m_eT} \pm p.
$$
It is straightforward to show (see Ref.~\cite{prd11} for detail)
that using the semiclassical approximation~(\ref{dsfWKB}) in
Eqs.~(\ref{d1sw}) and~(\ref{d1s}) one arrives at the FE
expressions~\cite{vogel89}
\begin{eqnarray}{d \sigma_{(w)}^{\rm FE} \over dT } &=& {G_F^{2}m_e \over 2 \pi} \left ( 1+ 4\sin^{2} \theta_{W} + 8 \sin^4 \theta_{W} \right
), \label{d1swFE}\\
{d \sigma_{(\mu)}^{\rm FE} \over dT } &=& \frac{4 \pi \alpha
\mu_{\nu}^{2}}{T}. \label{d1sFE}
\end{eqnarray}

Let us take into account the atomic recoil. This means that the energy
transfer $T$ is partly deposited in the center-of-mass atomic motion. The
energy of this motion, that is, the recoil energy, is $T_R=q^2/(2M_a)$,
where $M_a$ is the atomic mass. Hence, the first argument of the function
$S(T,{q}^{2})$ in~(\ref{d1sw}) and~(\ref{d1s}) must be replaced with
%\begin{equation}
%\label{dsf1} \theta(T_q-I)\,S(T_q,{q}^{2}),
%\end{equation}
%%
%
$$
T_q=T-T_R=T-\frac{q^2}{2M_a}.
$$
Substituting $T\to T_q$ in Eq.~(\ref{dsfWKB}),
% and neglecting the terms
%$\propto m_e/M_a$
we obtain
\begin{equation} S(T_q,{q}^{2})={m_e \over 2pq} \, \theta(q_a-q)
\, \theta(q_{+}-q)\,\theta(q-q_{-}), \label{dsfWKB1}
\end{equation}
where $q_a=\sqrt{2M_a(T-I)}$. From Eq.~(\ref{dsfWKB1}) it follows that the
atomic recoil has no effect if $q_a>q_{+}$. Indeed, in the latter case the
FE results~(\ref{d1swFE}) and~(\ref{d1sFE}) for the cross
sections~(\ref{d1sw}) and~(\ref{d1s}), respectively, remain unaltered. In
contrast, when $q_a<q_{+}$ in Eq.~(\ref{dsfWKB1}), the cross sections
become suppressed relative to the respective FE values and even vanish at
$q_a=q_{-}$.

Clearly, the dynamical structure factor of a real atom exhibits more
complex dependencies on $T$ and $q$ as compared to its semiclassical
approximation~(\ref{dsfWKB1}). Nevertheless, this approximation mimics the
main qualitative features pertinent to the exact function $S(T_q,q^2)$,
for example, such as ``spread and shift'' of the free-electron $\delta$
peak~\cite{prd11}. Thus, the criterion for the atomic-recoil effects to
come into play can be formulated for a real atom as $q_a\lesssim q_{+}$
or, accordingly, $T-I\lesssim 2p^2/M_a$. At such energy-transfer values
the cross sections~(\ref{d1sw}) and~(\ref{d1s}) become suppressed and
vanish when $T\to I$.

General numerical estimates can be obtained within the Thomas-Fermi model
of manyelectron atoms. The scale for the average electron momentum $p$ in
this model is determined by $\alpha Z^{2/3}m_e$. It gives the following
estimate for the energy range where the atomic-reoil effects are
important:
$$
T-I\lesssim2Z^{4/3}E_h(m_e/M_a),
$$
where $E_h=\alpha^2m_e=27.2$\,eV is the Hartree energy. For germanium
($Z=32$) we obtain $T-I\lesssim0.04$\,eV. This energy scale is
insignificant for the experiments searching for magnetic moments of
reactor antineutrinos~\cite{beda12}. As mentioned in the introduction,
these experiments have reached threshold values of $T$ as low as few keV.
Such values are already below the ionization threshold for $K$ electrons
in germanium ($I_K\approx11$\,keV). These atomic electrons are most
strongly bound, and the Thomas-Fermi average $p$ value substantially
underestimates their average momentum. The latter can be evaluated using
the virial theorem as $p=\sqrt{2m_eI_K}$. The corresponding energy scale
is $T-I\lesssim0.3$\,eV. It appears to be about an order of magnitude
larger than that given by the Thomas-Fermi model, but it is still
insignificant for the discussed experiments.

%The atomic recoil is usually neglected under the assumption
%$T\gg2E^2_\nu/M_a$, where $M_a$ is the nuclear mass.

%
\section{Conclusions}
\label{concl}
In summary, we have inspected how the center-of-mass atomic motion can
affect ionization of an atom by neutrino impact. Employing the
semiclassical approximation, we have derived the criterion that defines
the energy-transfer range where the atomic-recoil effects are important.
Our numerical estimates have shown that these effects play no appreciable
role in current experiments searching for neutrino magnetic moments of
electron antineutrinos from reactors with Ge detectors.

\section*{Acknowledgments}
One of the authors (A.S.) is thankful to Arcadi Santamar\'{\i}a, Salvador Mart\'{\i} and Juan Fuster for the kind invitation to participate at the ICHEP 2014 conference and to all of the organizers for their hospitality in Valencia. This work was supported by the Russian Foundation for Basic Research
(Grants no. 14-01-00420-a and no. 14-22-03043-ofi-m).



%% References

%\nocite{*}
%\bibliographystyle{elsarticle-num}
%\bibliography{martin}



\end{document}